\documentclass[a4paper,
              ]{jacow}
\makeatletter
\ifboolexpr{bool{xetex}}                 
 {\renewcommand{\Gin@extensions}{.pdf,%
                    .png,.jpg,.bmp,.pict,.tif,.psd,.mac,.sga,.tga,.gif,%
                    .eps,.ps,%
                    }}{}
\makeatother
\ifboolexpr{bool{xetex} or bool{luatex}} 
 {}                                      
 {\usepackage[utf8]{inputenc}}           
\usepackage[USenglish]{babel}
\ifboolexpr{bool{jacowbiblatex}}
 {%
  \addbibresource{jacow-test.bib}
  \addbibresource{biblatex-examples.bib}
 }{}
\usepackage{amsmath}

\pdfoutput=1
\begin{document}
\title{Non-interleaved FFS design}
\author{O. Blanco\thanks{LAL, Universite Paris-Sud, CNRS/IN2P3, Orsay, France}\thanks{CERN, Geneva, Switzerland}, R. Tomas$^\dag$, P. Bambade$^*$}
\maketitle
\begin{abstract}
The requirements on the Final Focusing System (FFS) for a new linear collider has lead to lattice designs where chromaticity is corrected either locally or non-locally. Here, we present the status of an alternative lattice design developed for the current CLIC 500GeV beam parameters where chromaticity is corrected upstream the Final Doublet (FD) and the geometrical components in the vertical plane are corrected locally. The diagnose of this design showed to possible variations two the lattice : the second order dispersion and its derivative with respect to $s$ must be cancelled before the FD, or some dispersion must be tolerated at the sextupole located inside the FD.
\end{abstract}
\section{INTRODUCTION}
The vertical beam size at the Interaction Point (IP) requires to be minimized to obtain the designed luminosity in linear accelerators \cite{Schulte}, and the FFS lattice is conceived to perform the beam demagnification in the linear and non-linear regime. One of the main second order components is the chromatic effect, it is the change on path length along the line as a function of energy, changing the focal point and diluting the beam size. Two different proposals to compensate this effect are : the non-local and the local correction.\par
\begin{figure}[!htb]
   \centering
   \includegraphics*[scale=0.13,angle=0]{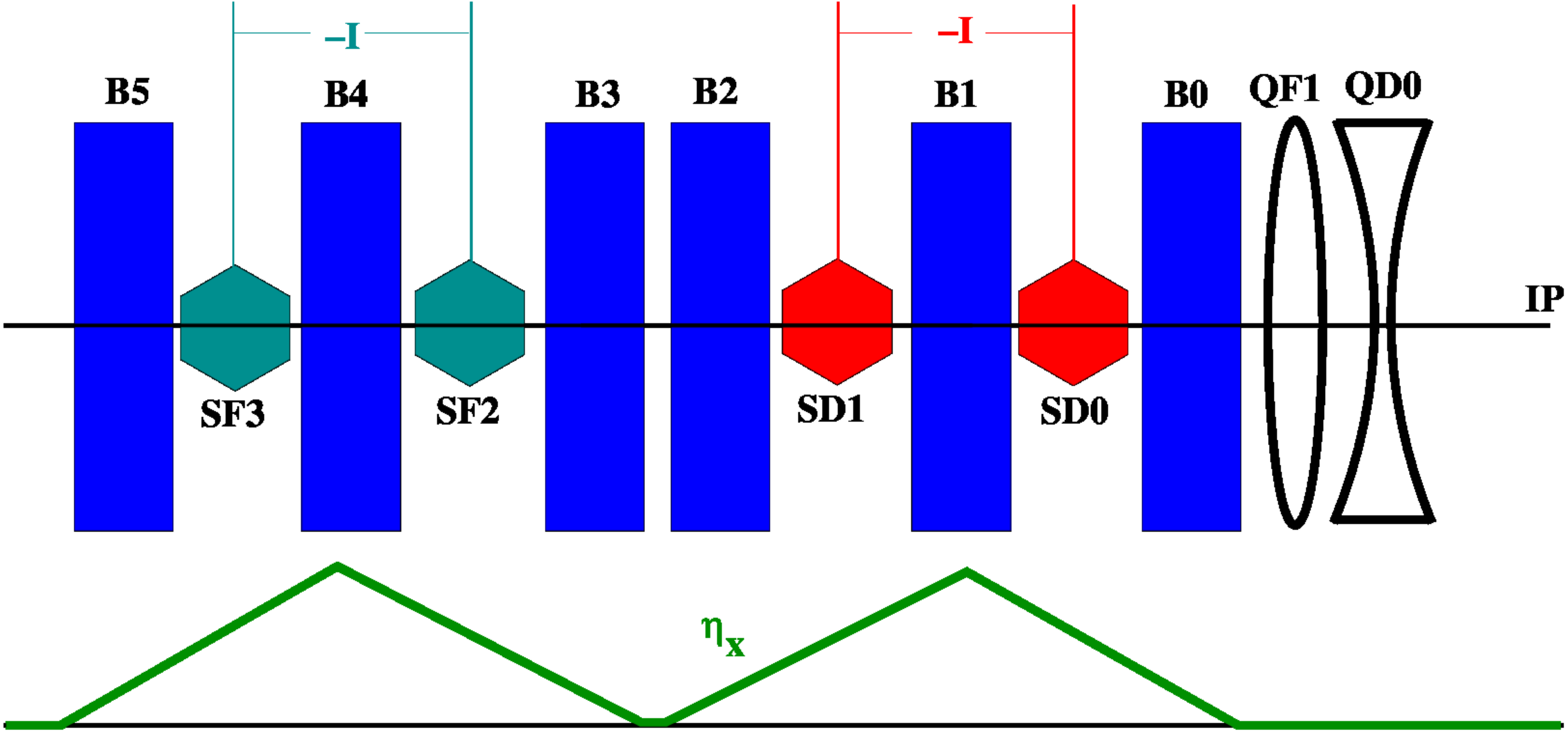}
   \caption{Non-local chromaticity correction.}
   \label{f-Non-local}
\end{figure}

The non-local correction scheme \cite{Brown-nl} compensates upstream the chromaticity in the Final Doublet (FD). A pair of sextupoles is used in horizontally dispersive regions ($\eta_x\neq0$) to compensate the path difference with respect to energy. In addition, they are matched to cancel geometrical aberrations, known as -I transformation. Fig. (\ref{f-Non-local}) shows schematically the lattice configuration, where QF1 and QD0 constitute the FD, B0 to B5 are dipole magnets to produce horizontal dispersion, SD and SF sextupoles are used to cancel vertical and horizontal chromaticity respectively.\par
\begin{figure}[!htb]
   \centering
   \includegraphics*[scale=0.13,angle=0]{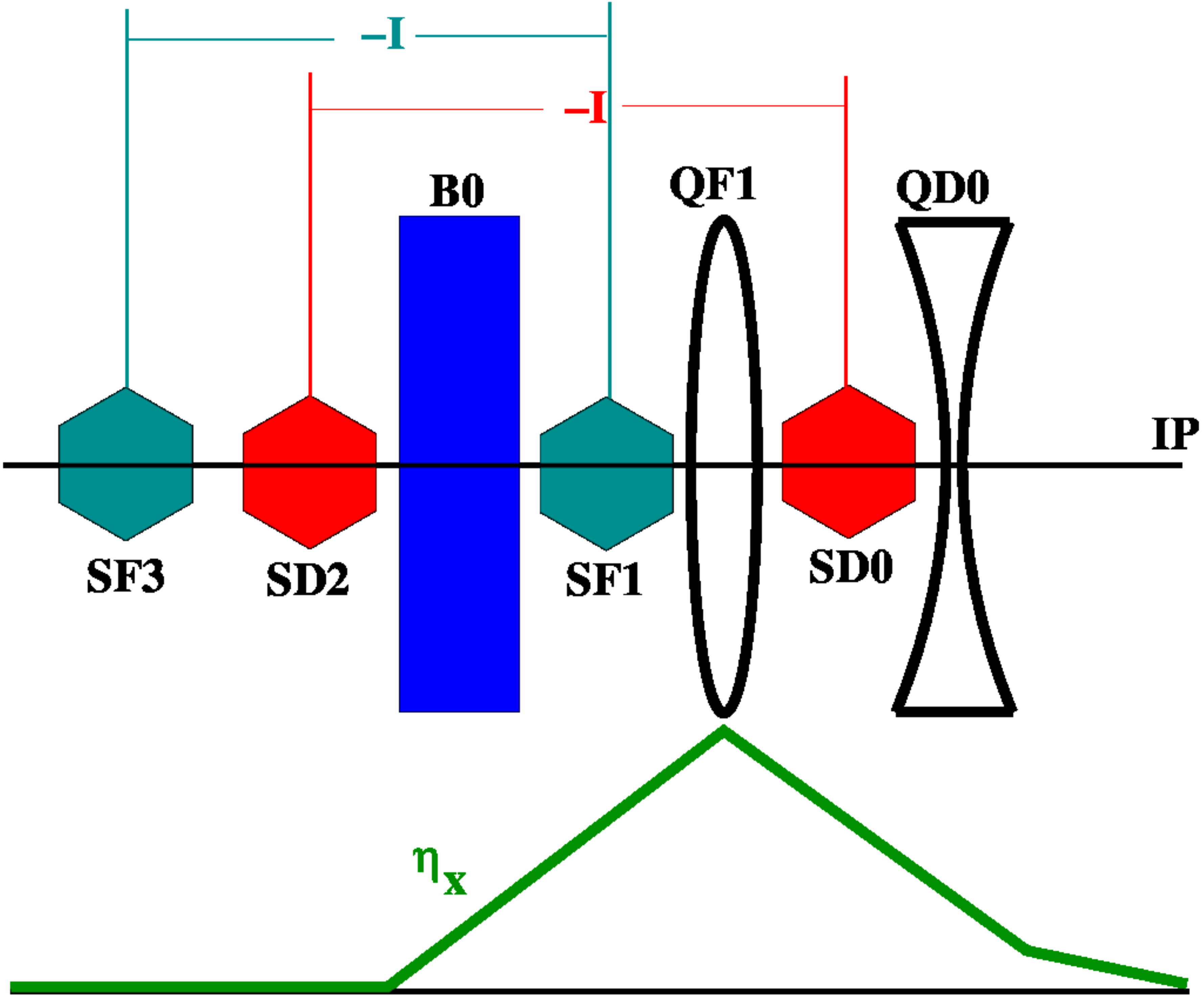}
   \caption{Local chromaticity correction.}
   \label{f-local}
\end{figure}

The local chromaticity correction scheme \cite{Raimondi-Seryi} compensates chromaticity inside the FD and Fig. (\ref{f-local}) shows the sextupoles locations. Here, one of the paired sextupoles is outside the horizontally dispersive region to cancel only geometrical contributions to beamsize. Horizontal dispersion is cancelled at the IP, and is generated only once upstream where the vertical and horizontal corrections share the region. This leads to a more compact design.\par
Both schemes have been compared for CLIC \cite{hgarcia}, concluding in smaller lattice length for the local scheme and easier tuning capabilities for the non-local. It has been mainly attributed to knobs orthogonality because the non-local scheme has a separated block per plane, while in the local, the horizontal and vertical correction blocks are interleaved.\par
\section{THE NON-INTERLEAVED LATTICE}
In the non-interleaved scheme, the idea is to preserve the separation in the vertical and horizontal chromatic corrections  while cancelling the geometrical components in the vertical plane at the FD. One of the paired sextupoles is inside a horizontally dispersive region while the other remains outside to cancel geometric contributions.\par
\begin{figure}[!htb]
   \centering
   \includegraphics*[scale=0.13,angle=0]{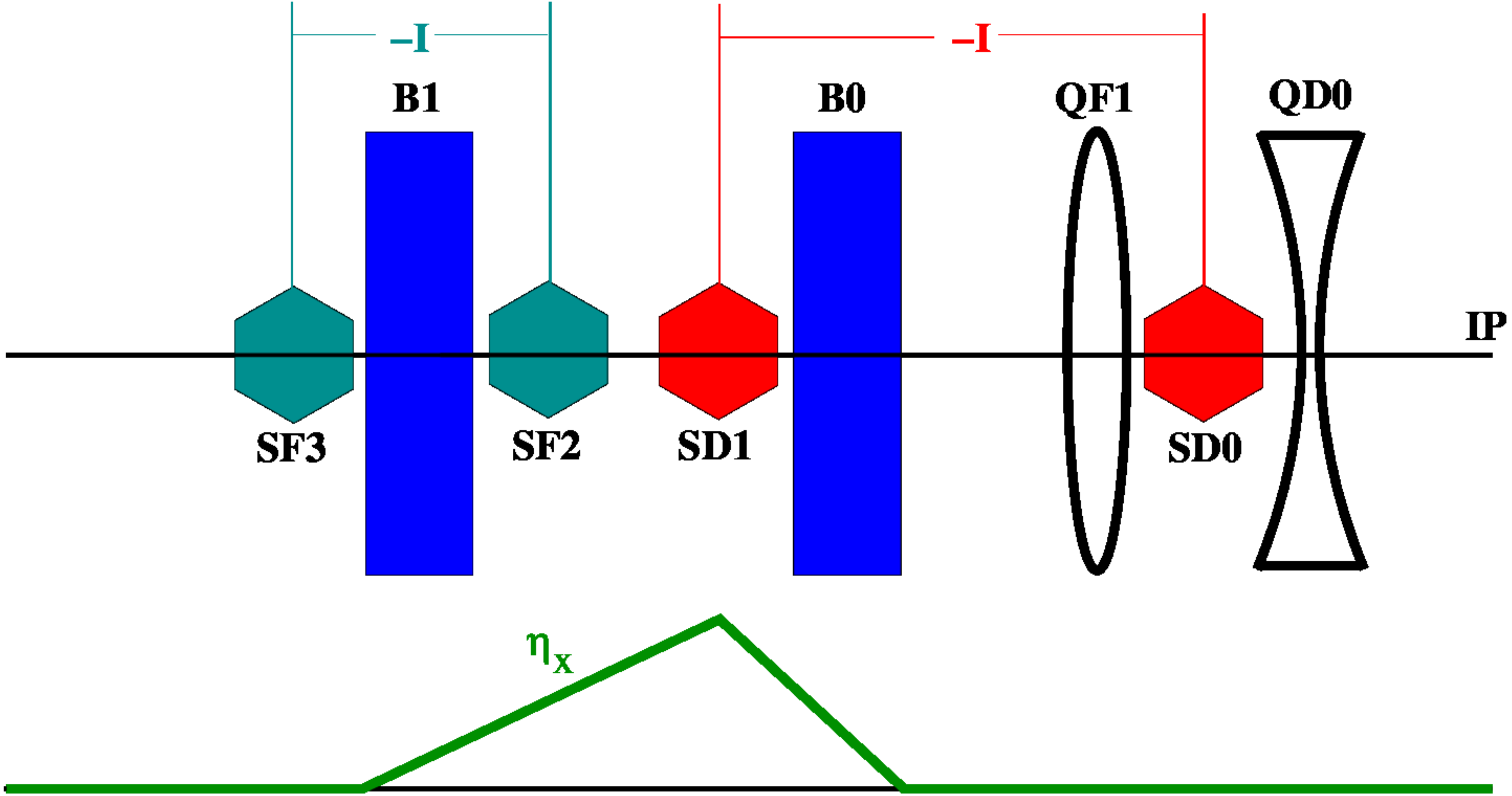}
   \caption{Non-interleaved chromaticity correction.}
   \label{f-noninterleaved}
\end{figure}
 Horizontal dispersion is generated over a common region and it is cancelled upstream, before the FD. Fig. (\ref{f-noninterleaved}) shows QD0 preceded by a sextupole SD0, which is matched with a second SD1 by the -I transformation. The same configuration is used to cancel horizontal chromaticity in an upstream section of the lattice, using SF sextupoles.\par
 
\section{CHROMATICITY MINIMIZATION}
\begin{figure}[!htb]
   \centering
   \includegraphics*[scale=0.35,angle=0]{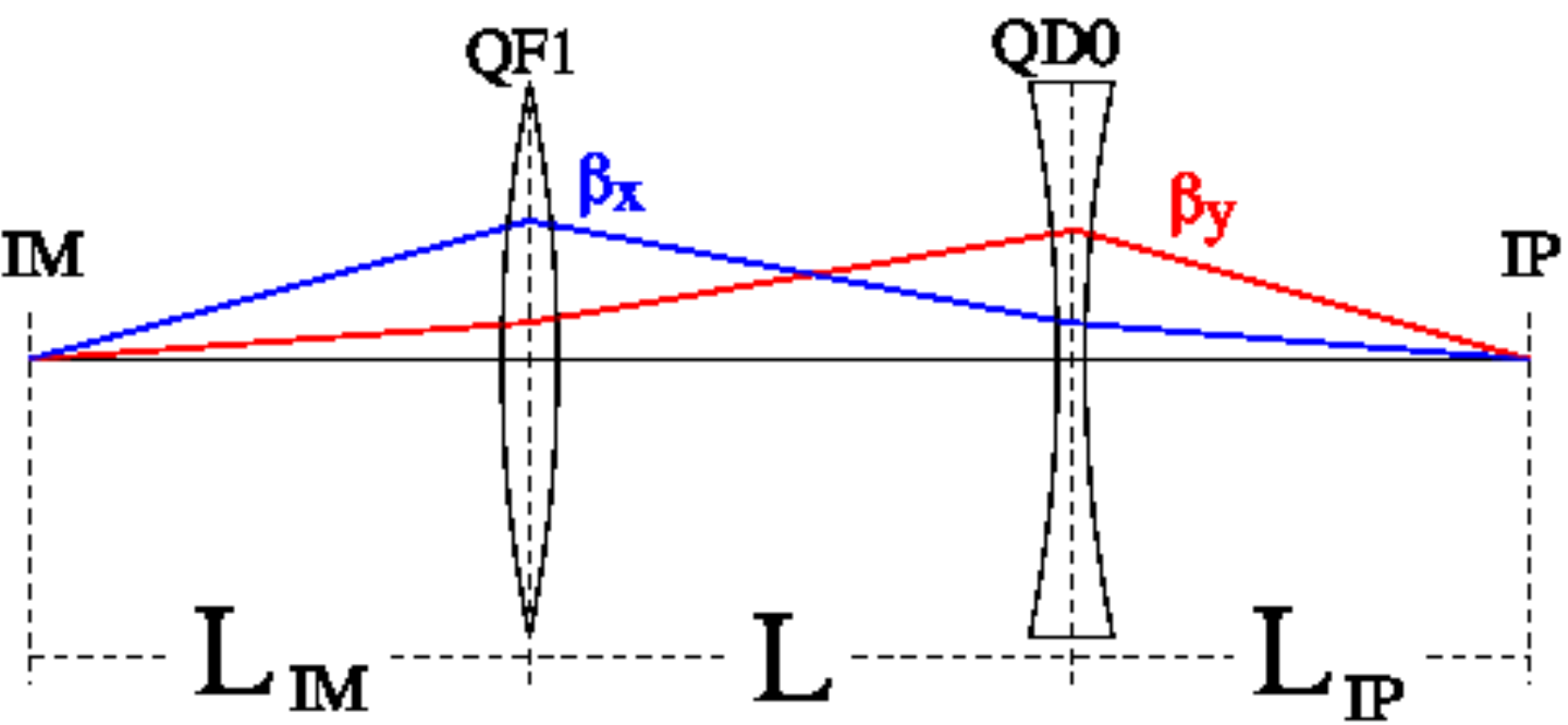}
   \caption{FD with focal points IM and IP at different distances.}
   \label{f-FD}
\end{figure}

Following the properties of telescopic systems \cite{Brown2}, a FD as in Fig. (\ref{f-FD}) with $r=L/L_{IP}$ and $r_{im}=L_{IM}/L_{IP}$, has peak $\beta$-functions at the quadrupoles equal to $\beta_{x0\atop y0}=\frac{L^2_{IP}}{\beta^*_{x\atop y}}$  for $QD0$ and $
\beta_{x1\atop y1}=\beta_{x0\atop y0}\left(1+r\pm\sqrt{\frac{r}{r_{im}}+r+\frac{r^2}{r_{im}}}\sqrt{\frac{1+r}{1+r/r_{im}}}\right)^2
$ for $QF1$, using $\left(\frac{\beta^*}{L_{IP}}\right)^2\ll1$.\par
In the thin lens approximation, chromaticity in the horizontal or vertical plane $\xi_{x\atop y}$ can be evaluated as\par	
$
 \xi_{x\atop y}=\mp\frac{1}{4\pi}\left(\beta_{x1\atop y1}k_1l_1-\beta_{x0\atop y0}k_0l_0\right)=\frac{1}{4\pi}\frac{L_{IP}}{\beta^*_{x\atop y}}\Xi_{x\atop y}(r,r_{im})
$, 
where $k,l$ are the quadrupole gradient and length, $\beta^*$ is the $\beta$ function at the IP and 
 \begin{align*}
 \Xi_{x\atop y}(r,r_{im})=&\mp\sqrt{\frac{1}{rr_{im}}+\frac{1}{r}+\frac{1}{r_{im}}}\sqrt{\frac{1+r/r_{im}}{1+r}}\\ &\left[\left(1+r\pm\sqrt{\frac{r}{r_{im}}+r+\frac{r^2}{r_{im}}}\sqrt{\frac{1+r}{1+r/r_{im}}}\right)^2\right.\\
 &\qquad\qquad\qquad\qquad\qquad\quad-\left.\left(\frac{1+r}{1+r/r_{im}}\right)\right]
\end{align*}\par
\begin{figure}[!htb]
   \centering
   \includegraphics*[scale=0.3,angle=-90]{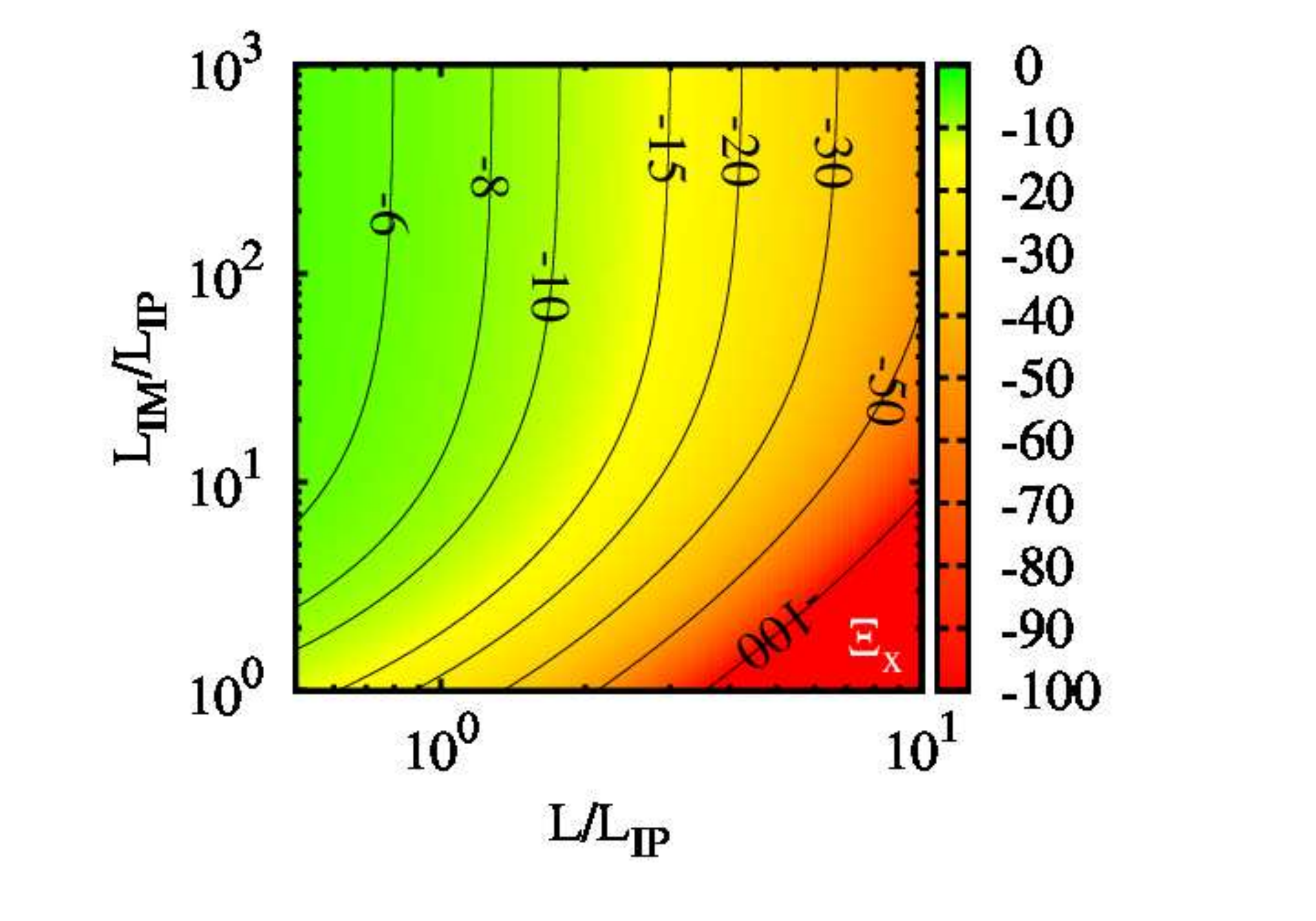}
   \caption{Horizontal normalized chromaticity.}
   \label{f-Xixn}
\end{figure}
\begin{figure}[!htb]
   \centering
   \includegraphics*[scale=0.3,angle=-90]{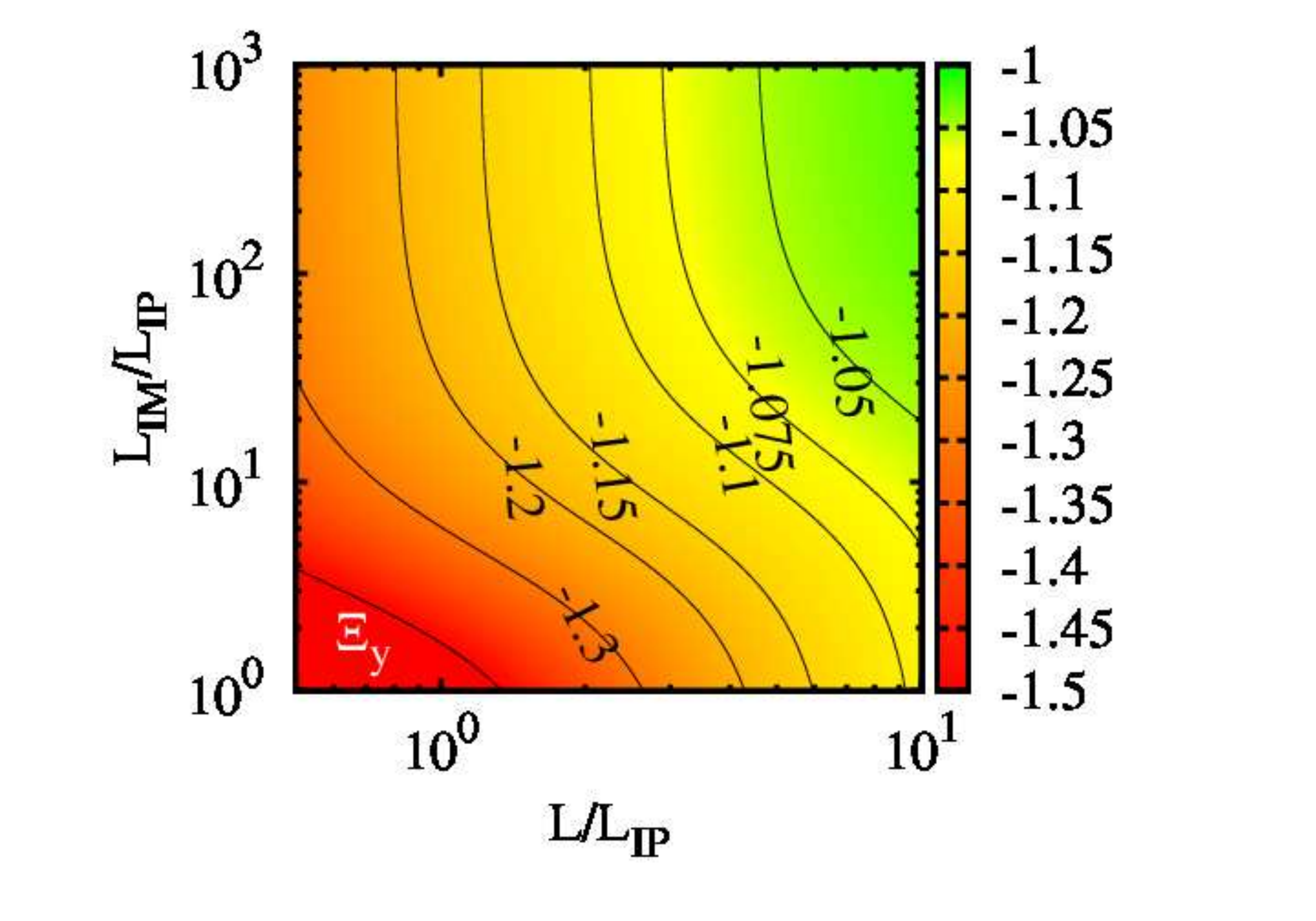}
   \caption{Vertical normalized chromaticity.}
   \label{f-Xiyn}
\end{figure}

Figures (\ref{f-Xixn}-\ref{f-Xiyn}) show the functions $\Xi_x$ and $\Xi_y$, equivalent to a normalized chromaticity in $\frac{1}{4\pi}L_{IP}/\beta^*$ units. The IP beam size is increased by chromaticity as $\sigma^*_{x\atop y}=\sigma_{{x0\atop y0}}\sqrt{1+\xi^2_{x\atop y}\sigma_\delta^2}$, where $\sigma_0^2=\beta^*\epsilon$, valid for both transversal planes. Therefore, luminosity is approximately affected by the horizontal and vertical chromaticity as \par
\begin{equation}
 L\propto\frac{1}{\sigma_x\sigma_y}=\frac{1}{\sigma_{x0}\sigma_{y0}\sqrt{1+(\xi^2_x+\xi^2_y)\sigma_\delta^2+\xi_x^4\xi_y^4\sigma_\delta^4}}
\end{equation}
\begin{figure}[!htb]
   \centering
   \includegraphics*[scale=0.3,angle=-90]{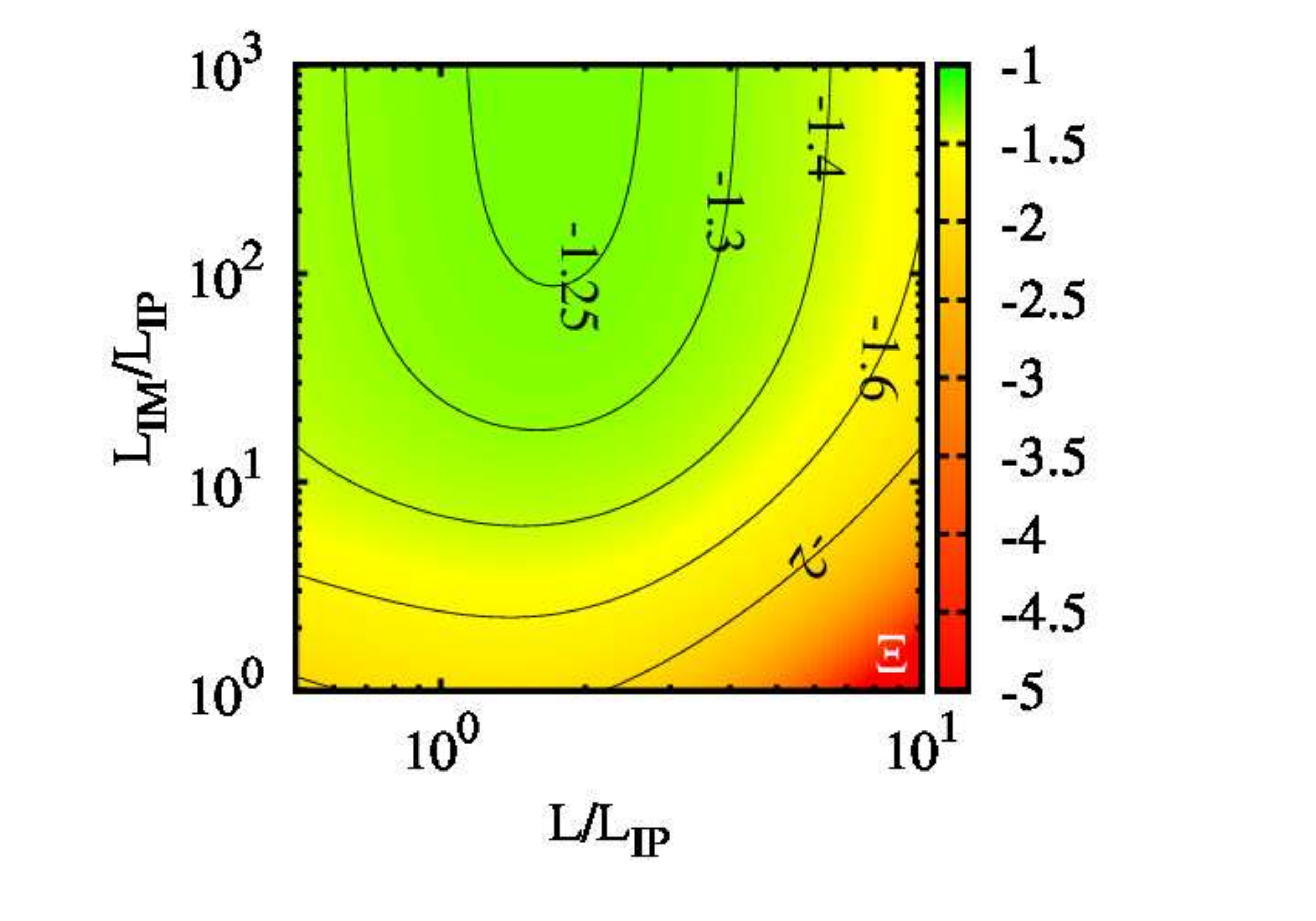}
   \caption{Added horizontal and vertical chromaticity for CLIC 500 GeV.}
   \label{f-Xixy500GeV}
\end{figure}
One possibility is to minimize the addition of the horizontal and vertical chromaticities,\par
\begin{align*}
 \xi = \xi_x + \xi_y &= \frac{1}{4\pi}\frac{L_{IP}}{\beta^*_y}\left(\frac{\Xi_x(r,r_{im})}{\beta^*_x/\beta^*_y}+\Xi_y(r,r_{im})\right)\\
 &= \frac{1}{4\pi}\frac{L_{IP}}{\beta^*_y}\Xi(\beta^*_x/\beta^*_y,r,r_{im})
\end{align*}
Figure (\ref{f-Xixy500GeV}) corresponds to $\xi$ for CLIC 500 GeV \cite{CLICdes}, $\beta^*_x/\beta^*_y=8.0\text{mm}/0.1\text{mm}=80$. For a given $L_{IM}$, the minimum added chromaticity in the FD is found when $L$ is one to two times the distance to the IP. In addition, Figures (\ref{f-Xixn}-\ref{f-Xixy500GeV}) show the effect of scaling up the system. For example, starting at $L=L_{IP}$ and $L_{IM}=10L_{IP}$, it is clear that a system with $L=10L_{IP}$ and $L_{IM}=100L_{IP}$, will decrease the vertical chromaticity by increasing the horizontal chromaticity.

\section{GEOMETRIC TERMS CANCELLATION}
\begin{figure}[!htb]
   \centering
   \includegraphics*[scale=0.15,angle=0]{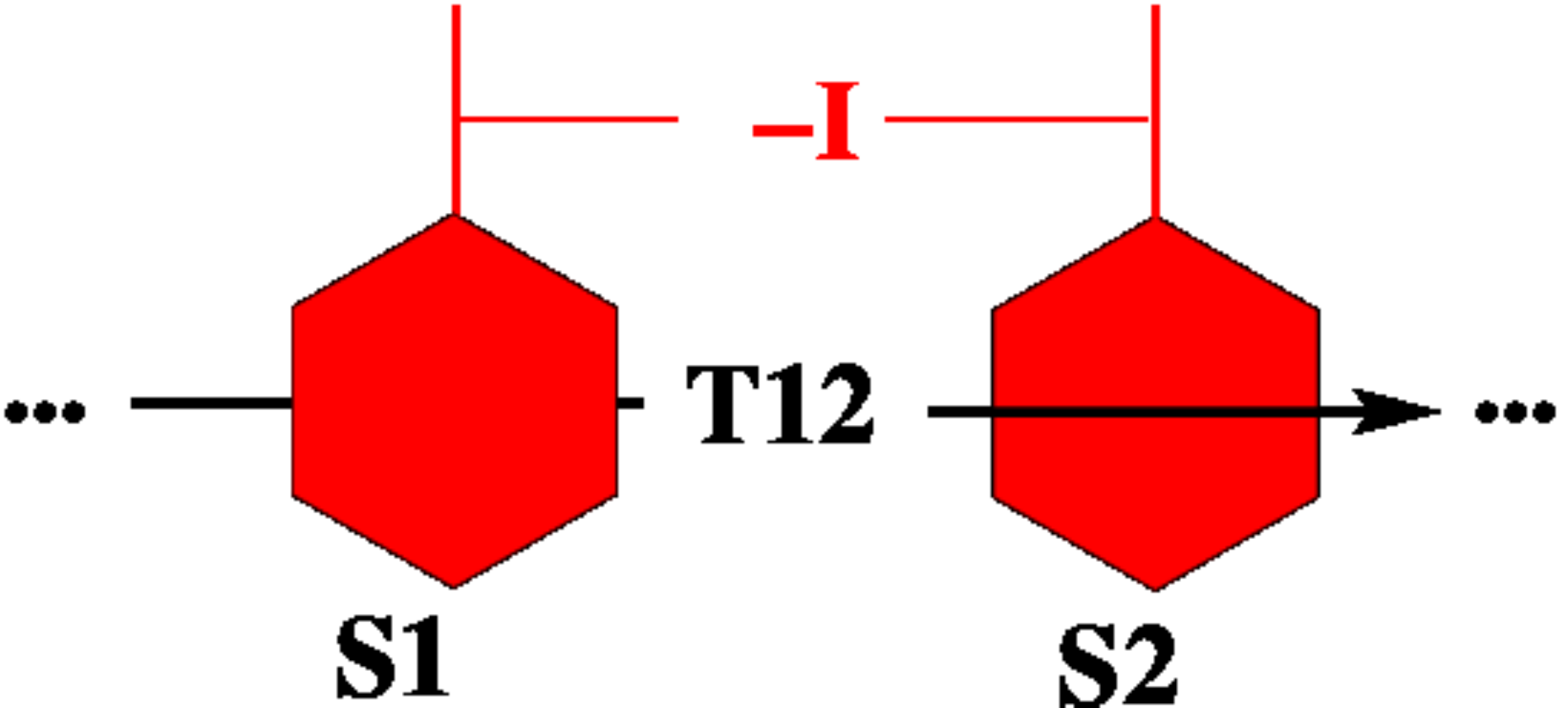}
   \caption{Sextupoles joined by the transport matrix $T_{12}$.}
   \label{f-sext}
\end{figure}

All chromatic correction schemes use a pair of matched sextupoles to cancel one another geometrical components. It is generally noted as the -I transformation. However, results from \cite{Xu} show that the -I transformation is one particular solution, and having a pair sextupoles as in Fig. (\ref{f-sext}) joined by a transport matrix $T_{12}$, the general solution is $t_{12}=0$, $t_{34}=0$, $t_{11}t_{22}=1$, $t_{33}t_{44}=1$, $k_{s1}=-k_{s2}t_{11}$ and $t_{11}=\pm t_{33}$, where, $t_{ij}$ are the matrix elements and $k_{s1}, k_{s2}$ are the sextupole gradients. This general solution is used here to give more flexibility to the beta functions in the design.\par
Those expressions from the linear transport map set restrictions on the optic lattice parameters and tolerances during the lattice design stage. In order to evaluate those tolerances, $\Delta\phi_x, \Delta\phi_y, r_x, r_y$ have been defined as $\phi_{x12}+\Delta\phi_{x} = \pi$, $\phi_{y12}+\Delta\phi_{y} = \pi$, $ k_{s1}\beta_{x1}^{3/2}-k_{s2}\beta_{x2}^{3/2}+r_x=0$ and $k_{s1}\beta_{y1}^{3/2}-k_{s2}\beta_{y2}^{3/2}+r_y=0$ respectively, where $\Delta\phi$ is the phase advance error, $r$ is the residual after subtractions and $\phi,\beta$ are the optical parameters in horizontal $x$ and vertical $y$ planes for both sextupoles $S1$ or $S2$.\par
Factors $\alpha_{x1}\Delta\phi_x$, $\alpha_{x2}\Delta\phi_x$, $\alpha_{y1}\Delta\phi_y$, $\alpha_{y2}\Delta\phi_y\ll1$, with $\alpha$ the optic parameter, are conditions to achieve geometrical terms cancellation. The FD requires phase advances finely matched because $\alpha$ and $\beta$ are high, and the residuals $r_x,r_y$ should be close to zero in order to cancel the second order map components in both planes at the same time. The $\beta_y/\beta_x$ ratio can be chosen to match sextupole strengths.\par

\section{THE LATTICE}
\begin{figure}[!htb]
   \centering
   \hspace*{-0.6cm}
   \includegraphics*[scale=0.34,angle=0]{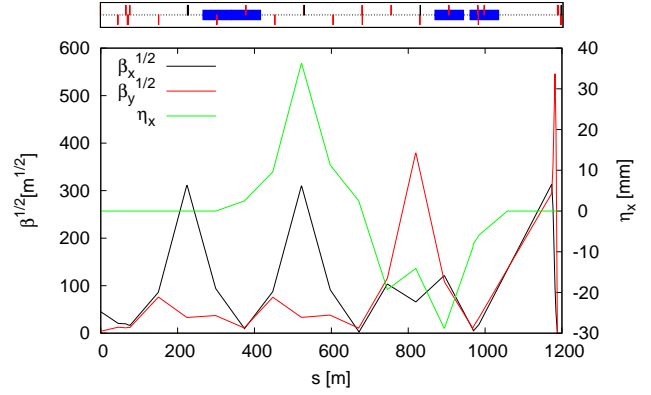}
   \caption{Non-interleaved lattice design for CLIC 500 GeV. Dipoles in blue, horizontally focusing quadrupoles in red and above the axis, vertically focusing quadrupoles in red and below the axis, and sextupoles in black.}
   \label{f-MAD-X}
\end{figure}

Using the CLIC 500 GeV parameters, new lattices were designed in MAD-X \cite{MAD-X} following the previous considerations and those in \cite{Seryi-Raimondi2}. Fig. (\ref{f-MAD-X}) is an example. It is worth noting the change of sign in the horizontal dispersion function $\eta_x$ from the horizontal chromatic correction to the vertical because this makes all sextupoles to have either possitive or negative gradients. Phase advances have been matched to $10^{-6}$ precision due to high $\alpha$ values in the FD. MAPCLASS2 \cite{Mapclassorig,Mapclass,Mapclass2,githubMapClass2} gives vertical beam size of 1.9nm and horizontal beam size of 186nm to the first order.\par
\begin{figure}[!htb]
   \centering
   \hspace*{-0.6cm}
   \includegraphics*[scale=0.34,angle=0]{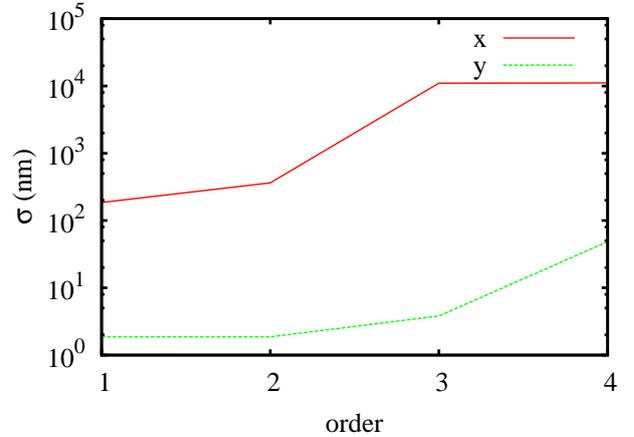}
   \caption{Beam size at the IP of the CLIC 500 GeV non-interleaved design as a function of transfer map order obtained with MAD-X PTC.}
   \label{f-beamsize}
\end{figure}

However, Fig. (\ref{f-beamsize}) shows that the horizontal beam size increases slightly to second order and by more than an order of magnitude when third order components in the map are considered.\par
\begin{figure}[!htb]
   \centering
   \hspace*{-0.6cm}
   \includegraphics*[scale=0.34,angle=0]{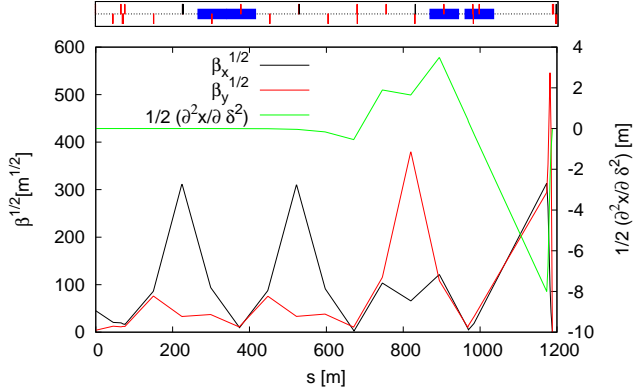}
   \caption{CLIC 500 GeV non-interleaved lattice. The second order dispersion from the $T_{166}$ map component is not zero at the sextupole in the FD.}
   \label{f-latticeT166}
\end{figure}

The reason to this beam size growth is the second order dispersion ($T_{166}$) in the sextupole inside the FD, see Fig. (\ref{f-latticeT166}). As opposed to the local chromaticity method were $T_{166}$ is cancelled only at the IP by matching the sextupoles and dispersion function, here, the second order dispersion generates higher order components due to the sextupole inside the FD used for geometrical cancellation. This is not present in the non-local method because there is no sextupole in the FD.\par
Two possible solutions are foreseen at the moment : cancel the map components $T_{166}$ and $T_{266}$ before the FD, or alternatively tolerate some dispersion in the FD to cancel the second order map and making it similar to the local method.\par
\section{CONCLUSION}
The non-interleaved lattice proposal has been conceived as an alternative to the local and non-local chromaticity correction methods. The added horizontal and vertical chromaticity has been minimized by using a distance from QD0 to QF1 approximately one and to times the distance between QD0 and the IP. Also, a general geometrical cancellation has been used to give more flexibility to the linear lattice design, however the large beta functions close to the FD impose high precision in the phase advance betweeen sextupole elements.\par
The non-interleaved design for CLIC 500 GeV has been diagnosed using MAD-X and MAPCLASS2, concluding that the second order dispersion must be cancelled before the FD because of the high gradient sextupole before QD0 used to cancel geometrical components only.\par
Two solutions are foreseen : the cancellation of the second order dispersion and its derivative before the FD, or alternatively generate dispersion in the FD to cancel the residual second and third order components.\par
%
\iffalse  
	\newpage
	\printbibliography

\begin{thebibliography}{99} 
\bibitem{Schulte} D. Schulte. Study of Electromagnetic and Hadronic Background in the Interaction Region of the TESLA Collider. Tesla-Report, 1997-08.
\bibitem{Brown-nl} K. Brown. Basic Optics for the SLC Final Focus. SLAC-PUB-4811, 1988.
\bibitem{Raimondi-Seryi} P. Raimondi, A. Seryi. A Novel Final Focus Design for Future Colliders. SLAC-PUB-8460. May, 2000.
\bibitem{hgarcia} H. Garcia, R. Tomas. Final-focus systems for multi-TeV linear colliders. Phys. Rev. ST Accel. Beams PhysRevSTAB.17.101001, Oct 2014.
\bibitem{Xu} G. Xu. General conditions for self-cancellation of geometric aberrations in a lattice structure. PRST-AB 8,104002, 2005.
\bibitem{Brown2} K. Brown. A Conceptual Design of Final Focus System for Linear Colliders. SLAC-PUB-4159. June 1987.
\bibitem{CLICdes}  Aicheler, M and Burrows, P and Draper, M and Garvey, T and Lebrun, P and Peach, K and Phinney, N and Schmickler, H and Schulte, D and Toge, N. A Multi-TeV Linear Collider Based on CLIC Technology: CLIC Conceptual Design Report. CERN-2012-007. SLAC-R-985. KEK-Report-2012-1. PSI-12-01. JAI-2012-001. Geneva, 2012.
\bibitem{MAD-X} MAD-X \url{http://mad.web.cern.ch/mad}\\Last visit 17 April 2013.
\bibitem{Seryi-Raimondi2} A. Seryi, M. Woodley, P. Raimondi. A Recipe for Linear Collider Final Focus System Design. SLAC-PUB-9895. May 2003.
\bibitem{Mapclassorig} R. Tomas. Nonlinear optimization of beamlines. PRST-AB 9, 081001 (2006).
\bibitem{Mapclass} R. Tomas. MAPCLASS: a code to optimize high order aberrations. CERN-AB-Note-2006-017. January 15, 2007.
\bibitem{Mapclass2} D. Martinez et al. MAPCLASS2: a code to aid the optimisation of lattice design. CERN-ATS-Note-2012-087 TECH. November 01, 2012.
\bibitem{githubMapClass2} MapClass2 \url{https://github.com/pylhc/MapClass2}\\Last visit 13 January 2014.
\end{thebibliography}
\else

\end{document}